\def\be{\begin{equation}}
\def\ee{\end{equation}}
\def\ber{\begin{eqnarray}}
\def\eer{\end{eqnarray}}
\def\bers{\begin{eqnarray*}}
\def\eers{\end{eqnarray*}}

\documentclass[aps,prapplied,twocolumn,showpacs,superscriptaddress, amsmath,amssymb,reprint,longbibliography]{revtex4-2}

\usepackage{dcolumn}   % Align table columns on decimal point
\usepackage{amsmath}
\usepackage{multirow}
\usepackage{graphicx}    % Include figure files
\usepackage{subfigure}
\usepackage{comment}
\usepackage{color}
\usepackage{makecell}
\usepackage{longtable}
\usepackage{natbib}
\usepackage{ifthen}
\usepackage{nicefrac}
\newboolean{includefigs}
\setboolean{includefigs}{true}      % Permits rapid compiling by removing figures/text
\newboolean{includetext}
\setboolean{includetext}{true}     % false = removed, true = included
\newcommand{\condcomment}[2]{\ifthenelse{#1}{#2}{}}
\begin{document}

%%%%%%%%%%%%%%%%%%%%%%%%%%%%%%%%%%%%%%%%%%%%%%%%%
%               TITLE
%%%%%%%%%%%%%%%%%%%%%%%%%%%%%%%%%%%%%%%%%%%%%%%%
\title{Carrier Selectivity and Passivation at the Group V elemental 2D Material--Si Interface of a PV Device}

\author{Gurudayal Behera}
\affiliation{Department of Energy Science and Engineering, IIT Bombay, Powai, Mumbai 400076, India}

\author{K. R. Balasubramaniam}
\email{bala.ramanathan@iitb.ac.in}
\affiliation{Department of Energy Science and Engineering, IIT Bombay, Powai, Mumbai 400076, India}

\author{Aftab Alam}
\email{aftab@iitb.ac.in}
\affiliation{Department of Physics, Indian Institute of Technology, Bombay, Powai, Mumbai 400076, India}

%%%%%%%%%%%%%%%%%%%%%%%%%%%%%%%%%%%%%%%%%%%%%%%%%
%               ABSTRACT
%%%%%%%%%%%%%%%%%%%%%%%%%%%%%%%%%%%%%%%%%%%%%%%%%
\begin{abstract}
This study investigates the interfacial characteristics relevant to photovoltaic (PV) devices of the Group--V elemental 2D layers with Si. The surface passivation and carrier selectivity of the interface between $\alpha$ and $\beta$ allotropes of arsenene, antimonene, and bismuthene monolayers with Si (100) and Si(111) were estimated \emph{via} first--principles calculations. Amongst the various interface configurations studied, all of the Si(111)--based slabs and only a couple of the Si(100)--based slabs are found to be stable. Bader charge analysis reveals that charge transfer from/to the Si slab to (As)/from (Sb and Bi) in the 2D layer occurs, indicating a strong interaction between atoms across the interface. Comparing within the various configurations of a particular charge (electron or hole) selective layer, the structural distortion of the Si slab is the lowest for $\alpha$--As/Si and $\beta$-Bi/Si. This translates as a lower surface density of states (DOS) in the band gap arising out of the Si slab when integrated with $\alpha$--arsenene and $\beta$--bismuthene, implying better surface passivation.  All-in-all, our analysis suggests $\alpha$-As as the best candidate for a passivating electron selective layer, while $\beta$-Bi can be a promising candidate for a passivating hole selective layer.
\end{abstract}
\date{\today}
\maketitle
%%%%%%%%%%%%%%%%%%%%%%%%%%%%%%%%%%%%%%%%%%%%%%%%%
\section{Introduction}
Recent years have witnessed an explosive interest on two-dimensional (2D) materials because of their fascinating optoelectronic and transport properties, for instance in spintronics and optoelectronic devices \cite{liu2016van,pumera20172d,xiao2012coupled,novoselov2004electric,zhang2018recent,durgun2005silicon,kim2012synthesis,cahangirov2010armchair,csahin2009monolayer,wang2018transport,ahn20202d,choudhuri2019recent,zhang2021recent,premasiri2019tuning,elahi2022review}. Graphene\cite{novoselov2004electric,neto2009electronic,peng2014new}, and more recently, many other new mono--elemental 2D materials have been utilized in various applications owing to their appropriate band gap, high electron/hole mobility, high conductivity, \emph{etc.} In particular, monolayer 2D crystals derived from the group--V elements \emph{i.e.}, As, Sb, and Bi are purported to be attractive candidates as charge transport layers in optoelectronic devices \cite{pumera20172d,zhang2018recent,jamdagni2018two,zhang2018first,zhang2015atomically,shah2020experimental,wu2020electrical,mardanya2016four}. Recent studies reveal that various monolayer allotropes of such group-V 2D elements are dynamically stable, with band gaps ranging between 1.0 eV to 2.5 eV and hole mobility $\sim$10$^4$ cm$^2$V$^{-1}$s$^{-1}$.  In addition, the reflectivity of these 2D structures is low, with almost negligible absorption in the visible region. Such a combination of optical and electrical properties makes them suitable as transparent conducting (TC) layers in photovoltaic (PV) applications \cite{behera2023two,xu2017first,singh2016antimonene,yuan2020recent,shu2018electronic}.  However, the applicability of these 2D structures for other inter--layers such as passivating layer, electron/hole selective layer, or a combination of both in solar PV devices needs further exploration.

Recent advancement in passivating and carrier selective contacts featuring dual functional layers at the semiconductor surface, which passivates the surface and extract only one type of charge carrier (hole or electron) is recognized as one of the key strategies to achieve high efficiency Si PV cells \cite{gerling2016transition,astha2017performance,gerling2017origin,patawardhan2019effect,liang2015interaction,bivour2015alternative,battaglia2014silicom,geissbuhler2015efficient,gerling2015characterization,yang2016high,nagamastu2015titanium,battaglia2014hole}. Materials used as passivating and hole transport layer in Si-based solar PV devices should possess certain desired characteristics \emph{e.g.,} wide band gap (3--3.3 eV), high transparency and high work function (typically between 5--7 eV) that can provide large conduction band offset. On the other hand, for the electron transport layer, the material should have a lower work function than Si, creating a large valence band offset and allowing only electrons to pass through. It has been reported that the most stable allotropes (\emph{e.g.,} $\alpha$ and $\beta$) of the group--V 2D As, Sb, and Bi monolayers are highly transparent and have work functions in the range between 4 eV to 5.5 eV.\cite{jamdagni2018two,behera2021first} This indicates that these 2D monolayer systems have the potential to serve as promising passivating and carrier selective contact to solar PV devices. Therefore, a feasibility study to evaluate the potential of these 2D materials as efficient surface passivation and/or carrier transport layers to solar PV devices needs to be carried out by creating a heterostructure between the 2D layered structures and Si absorber.

Matthieu \emph{et al.}\cite{fortin2017synthesis, fortin2018recovering} studied monolayer $\beta$-As/Si(111) and $\beta$-Sb/Ge(111) interfaces using a density functional theory (DFT) approach, where both surfaces of Si(111) and Sb(111) are passivated with hydrogen atoms. The electronic structures of these As(Sb)/Si(Ge) interfaces conforms that the 2D As or Sb monolayers in this structure are semiconducting. A strong interaction between As--Si and Sb--Ge bonds at the interface results in strong orbital hybridization between them, leading to the saturation of dangling bonds on the Si and Ge surfaces. They have also successfully fabricated Sb/Ge heterostructures, paving the way to integrate such 2D materials as passivating contacts to PV devices \cite{fortin2017synthesis}. Similarly, Akturk \emph{et al.}\cite{akturk2015single} have reported the electronic band structures of monolayer $\beta$-Sb/Ge interfaces without any hydrogen passivation on Ge(111). They modeled two interface structures of $\beta$-Sb/Ge, (1) a monolayer of $\beta$-Sb on top of a monolayer of Ge(111) and (2) a monolayer of $\beta$-Sb on top of a trilayer of Ge(111) surface. In both cases, the electronic band structures reveal that the materials of $\beta$-Sb/Ge heterostructures remain metallic. However, a significant change in the electronic properties of antimonene is observed due to the substrate effect. In particular, a high density of states for the Sb layer at the Fermi level is found in the case of monolayer Sb/Ge heterostructures. In contrast, a significant interaction with the growth of Sb on three layers of Ge(111) modifies its electronic band structure and preserves its pristine monolayer character. As such, the choice of an optimized substrate thickness is crucial to achieve the desired electronic/optical properties of the 2D material/Ge(Si) interfaces.

The scant literature on the growth of $\beta$-arsenene and antimonene on Si and Ge substrates, is only superseded by the notable absence of studies concerning the other group-V based 2D structures, \emph{e.g.}, $\beta$-bismuthene as well as the $\alpha$ allotropes of arsenene, antimonene, and bismuthene. Thus, a comprehensive understanding of the electronic structures of all the 2D allotropes of As, Sb, and Bi monolayers/Si(111) interface. This is essential to evaluate their efficacy for surface passivation and carrier selectivity in Si PV devices, wherein surface texturing is the norm, thereby exposing (111) surfaces to the overgrowth. Also, in As(Sb)/Si(Ge) interfaces, both Si(111) and Ge(111) surfaces are passivated with hydrogen atoms. Investigating the interface interaction when 2D structures are directly in contact with the top side of the Si surface where only the bottom surface is passivated with hydrogen remains unexplored. Additionally, no study has been reported for 2D material/Si(100) heterostructures, which could hold significance for microprocessor applications.
  
In this study, we present the structural and electronic aspects of group--V elemental 2D material/Si((111) and (100) planes) interfaces employing first--principles DFT simulation. We aim to offer a better understanding of the proffered surface passivation effect of these 2D structures to Si surface. A careful analysis of the structural reconstruction near the interface of Si and group--V atoms highlights the important changes in bond lengths, and orbital hybridization and reveals their impacts on the optoelectronic properties. A strong interaction between Si and Group--V atoms located at/near the interface is observed \emph{via} Bader charge analysis; As atoms accept charge from Si while Sb and Bi donate charge to the Si atoms across the interface. The $\beta$--Bi/Si interface is less distorted than the more studied structures based on $\beta$ allotropes of As and Sb monolayers \cite{fortin2017synthesis, fortin2018recovering}, and can therefore be envisaged as a hole-selective layer with better surface passivation of Si. Importantly, we find the more critical passivating electron selective layer can be obtained with $\alpha$--As. The $\alpha$--As/Si interface exhibits better characteristics than the hole selective layers undertaken in this study and even some of the standard materials used as electron selective layers to Si \cite{gerling2016transition,astha2017performance,gerling2017origin,patawardhan2019effect,liang2015interaction,behera2021first}.\\
\section{Computational Details}
First--principles calculations were performed using density functional theory (DFT)\cite{Kohn}, as implemented within the Vienna Ab--initio Simulation Package (VASP) \cite{vasp,kresse1993ab,kresse1996efficient} with a projector augmented wave (PAW)\cite{paw} basis set. The generalized gradient approximation (GGA) \cite{perdew1997generalized} exchange--correlation functional as parameterized by the Perdew--Burke--Ernzerhof (PBE) potential \cite{perdew1996generalized} is used. The kinetic energy cut--off for the plane wave basis set was taken to be 550 eV, and the van der Waals (vdW) interactions are incorporated in all the calculations using the DFT--D2 method.\cite{grimme2006semiempirical} All the atoms in the unit cell are fully relaxed using the conjugate gradient method until the force (energy) converges below 0.001 eV/\AA$^2$\ (10$^{-6}$ eV). During structural optimization, the bottom layer of Si atoms and their corresponding passivated atoms are kept fixed to mimic the bulk nature of Si. The Brillouin zone (BZ) integration was done using a $\Gamma$--centered k-mesh scheme employing the tetrahedron method with Bl{\"o}ch corrections \cite{bolch}, as implemented in VASP. For a better understanding of charge distribution/transfer and the nature of bonding at the surface/interface, Bader charge analysis \cite{Henkel} was carried out. Due to the large supercell size of the heterostructures, the relaxation was carried out using 3$\times$6$\times$1 and 3$\times$3$\times$1 $\Gamma$--centered k--mesh sampling for $\alpha$ and $\beta$ allotropes heterostructures, respectively. However, to further simulate the electronic structure results such as density of states (DOS), Bader charge, surface formation energy, \emph{etc.} which are the essential quantities to assess the effect of passivation and charge transfer, a higher k--mesh of 6$\times$12$\times$1 and 12$\times$12$\times$1 were used for the self-consistent-field (SCF) calculations, respectively. The interface formation energy was calculated using the following expression,
\begin{equation}
	E_\mathrm{form}=\frac{E_\mathrm{Tot}-[nE_\mathrm{As/Sb/Bi}+mE_\mathrm{Si}+lE_\mathrm{H}]}{A}
\end{equation}
where, $\textquoteleft E_\mathrm{Tot}$\textquoteright\ is the total energy of the heterostructure unit cell, $\textquoteleft n$\textquoteright,\ $\textquoteleft m$\textquoteright,\ $\textquoteleft l$\textquoteright\ are the number of As/Sb/Bi, Si, and H atoms present in the cell. $E_\mathrm{As/Sb/Bi}$, $E_\mathrm{Si}$ and $E_\mathrm{H}$ are the total energy per atom of the respective ground state bulk structures of As/Sb/Bi, Si and H.
\begin{table*}[!]%
	\centering
	\caption{\label{tab:table1}% 
		Lattice mismatch, lattice parameters, surface formation energy ({\bf{E$_\mathrm{form}$}}), and the electronic band nature of the group–V 2D structures with Si(111) and Si(100) interfacial surface slabs. E$_\mathrm{g}$ represents the band gap. }
	%\hspace*{-3em}
	\begin{ruledtabular}
		\begin{tabular}{ccccc}
			\bf{Heterostructures}&\bf{Lattice} &\bf {Lattice parameters} (\AA)& \bf{E$_\mathrm{form}$}& \bf{Nature of}  \\
			\bf{type}&\bf{mismatch (\%)}&&\bf{(eV/\AA$^2$)}&\bf{Interfacial slabs}\\
			\hline \\
			$\alpha$--arsenene/Si(111):H& 3.5& a = 15.35, b= 7.67, c= 24.71 &0.04 &metallic\\
			$\beta$--arsenene/Si(111):H &4.5&a = 7.67, c= 22.35 &0.03&metallic\\
			$\beta$--antimonene/Si(111):H &1.4& a = 7.67, c= 22.66&0.03&metallic\\
			$\beta$--bismuthene/Si(111):H &4.5& a= 7.67, c=22.76&0.02&metallic\\
			$\alpha$--arsenene/Si(100):H &	4.8 &	a = 10.96, b= 5.07, c= 14.05&0.17&metallic\\
			$\alpha$--bismuthene/Si(100):H&2.9 &a= 10.86, c=16.42&0.13&E$_\mathrm{g}$= 0.05 eV\\
			$\beta$--arsenene/Si(100):H&4.5 &a= 10.86, b= 5.43, c= 11.68&0.14&E$_\mathrm{g}$= 0.42 eV\\
			$\beta$--antimonene/Si(100):H& 2.7 &a= 10.86, c=12.01&0.14&metallic\\
			$\beta$--bismuthene/Si(100):H&4.6 &a= 15.35, b= 10.86, c= 12.09&0.14&metallic\\
		\end{tabular}
	\end{ruledtabular}
\end{table*}

\section{Results and Discussion}
\subsection{Crystal Structures and Electronic Properties of Monolayer 2D-Structures/Si Interface}
The interfacial surface slabs of $\alpha$ and $\beta$ allotropes of the group-V 2D monolayers with six layers of Si(100) ($\sim$16.46 \AA\ thickness) and Si(111) ($\sim$8 \AA\ thickness) surfaces are modeled with a vacuum of 15 \AA. In order to obtain minimal lattice mismatch between $\alpha$ and $\beta$ allotropes of As, Sb, and Bi and Si(100) and Si(111) substrate, a 2$\times$1 supercell of Si surface(for both (100) and (111)) are considered. Therefore, the [100] directions of both 3$\times$2 supercells of $\alpha$-As, and 2$\times$2 supercell of $\beta$-As/Sb/Bi align with [100] direction of Si(111) surface. Similarly, [100] directions of 3$\times$2 $\alpha$-As and $\beta$-As/Sb/Bi supercell align with [100] directions of 2$\times$1 supercell of Si(100) substrate. The resulting interfacial surface slabs, which provide a minimal lattice mismatch between Si and group--V 2D monolayer allotropes, can be represented by the following epitaxial relationships:
\begin{itemize}
	\item (111)$_\mathrm{Si}$$\parallel$(110)$_\mathrm{As}$$\vert$2$\times$1[100]$_\mathrm{Si}$$\parallel$3$\times$2[100]$_\mathrm{\alpha-As}$$\vert$
	\item (111)$_\mathrm{Si}$$\parallel$(111)$_\mathrm{As/Sb/Bi}$$\vert$2$\times$1[100]$_\mathrm{Si}$$\parallel$2$\times$2[100]$_\mathrm{\beta-As/Sb/Bi}$$\vert$ 
	\item(100)$_\mathrm{Si}$$\parallel$(110)$_\mathrm{As}$$\vert$2$\times$1[100]$_\mathrm{Si}$$\parallel$3$\times$2[100]$_\mathrm{\alpha-As}$$\vert$
	\item(100)$_\mathrm{Si}$$\parallel$(111)$_\mathrm{As/Sb/Bi}$$\vert$2$\times$1[100]$_\mathrm{Si}$$\parallel$3$\times$2[100]$_\mathrm{\beta-As/Sb/Bi}$$\vert$
\end{itemize}
The atoms of the bottom Si layers were passivated with hydrogen atoms to avoid the effects of Si dangling bonds. This passivation confirms that the bulk--like bottom layer of Si does not significantly contribute to the surface states, and all the calculated gap states originate from the interfacial region. Table~\ref{tab:table1} display the lattice mismatch (in \%), surface formation energy ({\bf{E$_\mathrm{form}$}}), and lattice parameters of various 2D structures/Si interfaces. The interface slabs of $\alpha$-Sb/Si(111), $\alpha$-Bi/Si(111), and $\alpha$-Sb/Si(100) are not considered in the present study due to a large lattice mismatch (more than 7\%) between them. The surface formation energies listed in Table~\ref{tab:table1} show that all the group-V elemental 2D material/Si(111) interface structures are more stable than the 2D material/Si(100) interfaces (energy difference lying in the range $\sim$100--130 meV/\AA$^2$). All the 2D material and Si in various interface configurations show metallic features except $\alpha$-Bi/Si(100) and $\beta$-As/Si(100). From now onwards, we will mainly focus on the more stable 2D material/Si(111) heterostructures. For some instances, we shall also compare the electronic properties of $\alpha$-Bi/Si(100) and $\beta$-As/Si(100) surface slabs due to their interesting/contrasting electronic properties.

\begin{figure*}[t!]
	\centering
	\includegraphics[scale=0.55]{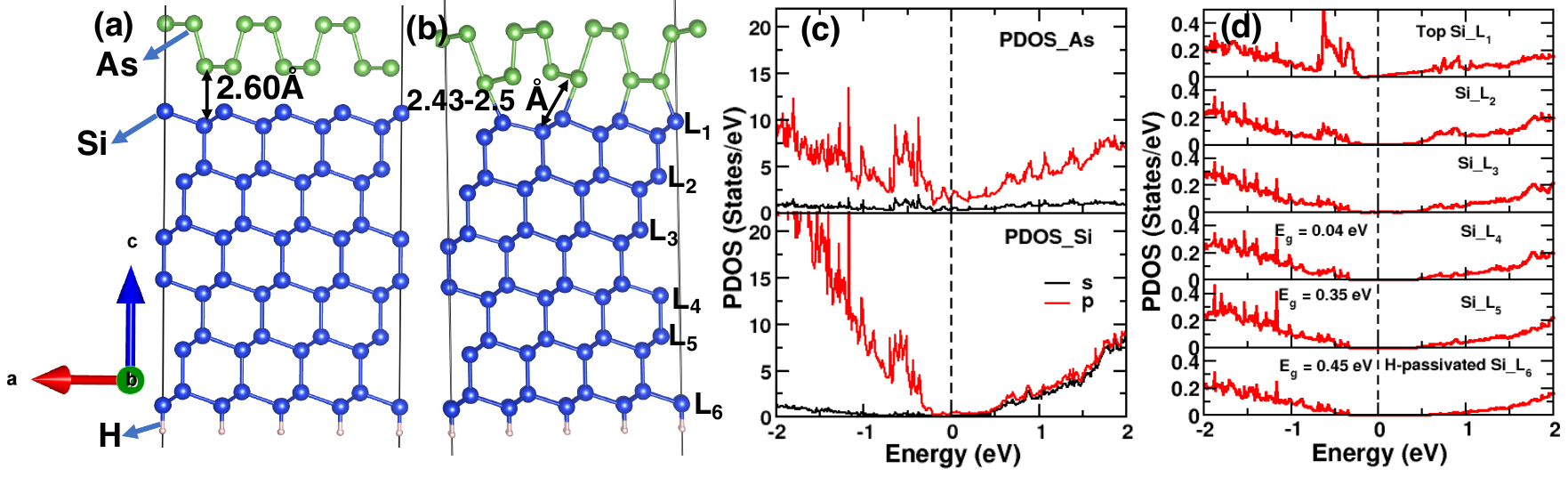}
	\caption{$\alpha$--As/Si(111):H interfacial structure (a) before and (b) after ionic relaxation. White, green, and blue spheres represent H, As, and Si atoms respectively. (c) Partial density of states (PDOS) of As and Si atom located at/near the interface and  (d) PDOS for various layers of Si atoms starting from the top (at/near interface) to bottom (H-passivated) surface layer. E$_\mathrm{f}$ is the Fermi level set at zero energy for each PDOS plot. }
	\label{fig1}
\end{figure*}

 Various possible configurations of monolayer $\alpha$--As/Si(111) interfaces are modeled by translating the As atoms with respect to the position of the top layer Si atoms to obtain the most stable heterostructures. The modeled structures and their corresponding surface formation energies of $\alpha$--As/Si interfaces are given in Fig.~S1 and Table~S1 of supplementary information (SI),\cite{supplement} respectively. The energetically most stable structure (before and after ionic relaxation) and the corresponding orbital projected density of states (PDOS) of relaxed structure are shown in Fig.~\ref{fig1}. After structural relaxation, As atoms interact with the Si atoms present at the top layer of the Si surface, forming different Si--As chemical bonds with bond lengths ranging between 2.43 \AA\ to 2.5 \AA. However, the Si--Si bond lengths remain 2.35 \AA\ (bulk--like) throughout the six layers of Si surfaces.

The orbital PDOS shows that the $\alpha$-As and Si layers at the interface are metallic in nature.  The metallicity in the $\alpha$--As/Si interface arises due to the significant contributions from the electronic states of As atoms (particularly p-orbitals) at the Fermi level (E$_\mathrm{f}$). However, very small contributions from Si atoms are also observed near E$_\mathrm{f}$, leading to rapid surface reconstruction on Si surface. The contribution of electronic states from each layer of Si surface (top to bottom) is plotted in Fig.~\ref{fig1}(d). The electronic states at/near E$_\mathrm{f}$ for each layer of Si atoms decrease as we move from top to bottom in the Si slab. A band gap value of 0.04 eV is found in the fourth layer, which increases further as we move towards the bottom layer. The band gap in the bottom layer of Si was found to be  0.45 eV, which behaves bulk--like Si. This indicates rapid surface reconstruction of the Si slab when interfaced with $\alpha$-arsenene monolayer.
\begin{figure*}
	\centering
	\includegraphics[scale=0.62]{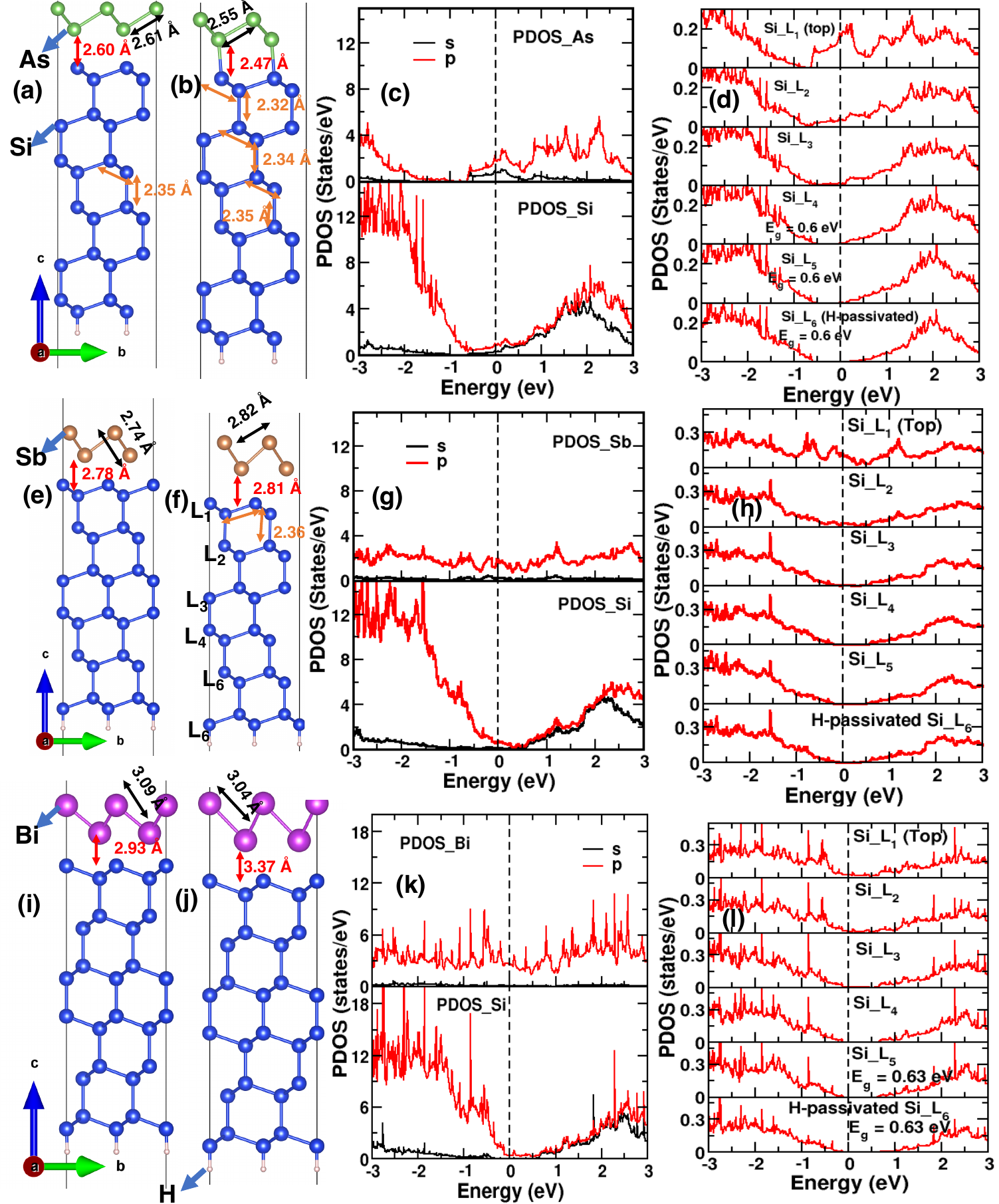}
	\caption{Heterostructures of $\beta$-As/Si(111):H, $\beta$-Sb/Si(111):H, and $\beta$-Bi/Si(111):H (a, e, i) before ionic relaxation (b, f, j) after ionic relaxation respectively. White, green, brown, magenta and blue spheres represent H, As, Sb, Bi, and  Si atoms, respectively. (c, g, k) Orbital projected DOS for s--orbital (black color) and p--orbital (red color) of As, Sb, Bi and Si atoms. (d, h, l) PDOS for various layers of Si atoms starting from the top (at/near interface) to bottom (H-passivated) Si-layer for $\beta$-As/Si(111):H, $\beta$-Sb/Si(111):H, and $\beta$-Bi/Si(111):H interfaces respectively. E$_\mathrm{f}$ is set at zero energy for each PDOS plot. }
	\label{fig2}
\end{figure*}

Similar to $\alpha$--As/Si interfaces, we have also modeled various possibilities of $\beta$ allotropes of 2D structure/Si(111) interfaces structures by translating the As/Sb/Bi atoms with respect to the position of the top layer Si atoms. All the possible modeled interfacial structures of $\beta$-As/Si, $\beta$-Sb/Si, and $\beta$-Bi/Si interfaces and their corresponding formation energies are shown in Fig.~S2 and Table~S2, respectively in SI.\cite{supplement} The most stable structures of all three interfaces (before and after ionic relaxation), and the corresponding PDOSs of relaxed structures are shown in Fig.~\ref{fig2}. After structural relaxation, the interfacial Si atoms closely interact with As atoms, forming a chemical bond of bond length 2.47 \AA\, as shown in Fig.~\ref{fig2}(b). Also, the structural distortion of the Si layer at the interface is observed. In contrast, the antimonene and bismuthene monolayers were found to interact extremely weakly with the interfacial Si layer post structural relaxation. In fact, they move away from the interfacial Si layer, maintaining a vertical layer to layer distance of 2.81 \AA\ and 3.37 \AA\ in case of $\beta$-Sb/Si(111) and $\beta$-Bi/Si(111), respectively (see Fig.~\ref{fig2}(f) and ~\ref{fig2}(j)). All three  2D monolayers themselves are found to distort slightly after structural relaxation. The bond length between As--As increases by 0.05 \AA\, while Sb--Sb bond length increases by 0.12 \AA\  with respect to the bond lengths before relaxation. However, the bond length of B-Bi in $\beta$-Bi/Si(111) surface slab maintains its initial value due to extremely weak interaction between Si and Bi atoms at the interface. Similarly, the Si--Si bond length in Si slab is found to change from 2.32 \AA\ (near the interface) to 2.36 \AA near the bottom bulk like Si--layers in the case of $\beta$-As(Sb)/Si(111) slabs. However, the Si--Si bond lengths in $\beta$-Bi/Si surface slabs remain 2.36 \AA (same as the bond length in bulk Si) in the entire Si slab (see Fig.~\ref{fig2}(j)), and a very minimal distortion in Si layers at the interface indicating minimal reconstructions in the Si surfaces.

Figure~\ref{fig2}(c, g, and k) display the orbital PDOS for $\beta$-As/Si(111), $\beta$-Sb/Si(111) and $\beta$-Bi/Si(111) interfaces. The DOS plot reveals that all the 2D materials at the interface structures are metallic in nature. The finite number of electronic states at E$_\mathrm{f}$ have significant contributions from p-orbitals of As, Sb, and Bi interfacial atoms as compared to the p-orbitals of interfacial Si atoms. To check the individual contribution of each Si--layer, we plotted the atom projected DOS of Si-atoms starting from top (interfacial) to the bottom (H-passivated) Si layer, which is shown in Fig.~\ref{fig2}(d, h and i). The figure clearly indicates that there is a large contribution of electronic states arising from Si-atoms, located at/near the interface in case of $\beta$-As(Sb)/Si(111) slabs compared to $\beta$--Bi/Si case. As we move away from the interface towards the bulk like bottom Si-layers (layer-1 to layer-6), the contribution of electronic states at E$_\mathrm{f}$ decreases, and a finite band gap of $\sim$0.6 eV is observed after three layers of Si surface. This indicates rapid surface reconstruction occurs in the Si slabs when interfaced with group-V elemental 2D monolayers.

As mentioned earlier, we have also studied the electronic structures for all the meta stable group--V 2D monolayers and Si(100) interfacial surface slabs. The detailed optimized interface structures and electronic properties of these 2D structure/Si(100) interfaces are discussed in Sec. III of SI\cite{supplement}. Interstingly, the orbital projected DOS of $\alpha$--Bi/Si(100) and $\beta$--As/Si(100) interfaces reveal that both the 2D monolayers preserving their semiconducting nature, with a band gap of 0.05 eV and 0.42 eV, respectively. The semiconducting nature suggests that $\alpha$-bismuthene and $\beta$-arsenene monolayers can effectively saturate the Si dangling bonds and mimic the verisimilitude bulk character of the Si surface. Therefore, it is reasonable to speculate that $\alpha$-bismuthene and $\beta$-arsenene can also be potential candidates for passivating layer to Si surface in solar PV devices.
\begin{figure*}[htbp!]
	\centering
	\includegraphics[scale=0.49]{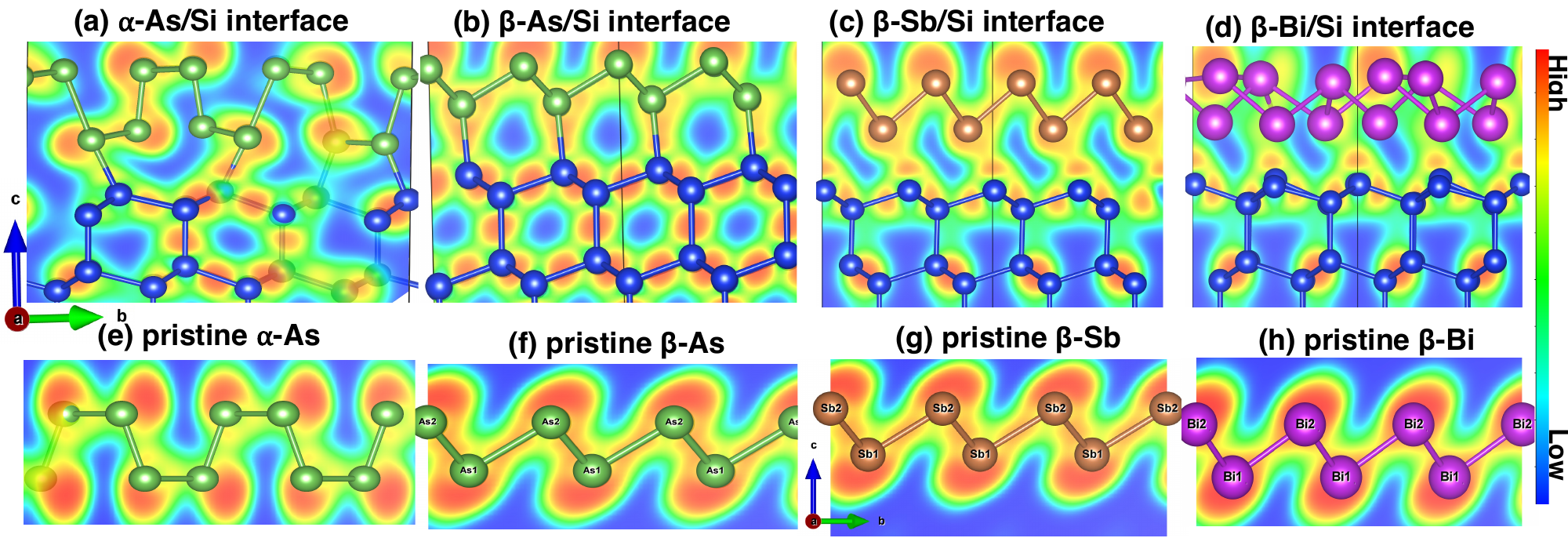}
	\caption{Charge density profile of (a) $\alpha$--As/Si(111):H, (b) $\beta$--As/Si(111):H, (c) $\beta$--Sb/Si(111):H, and (d) $\beta$--Bi/Si(111):H interface structures. (e, f, g, h) charge density profile of pristine $\alpha$--arsenene, $\beta$--arsenene, antimonene, and bismuthene, respectively. }
	\label{fig4}
\end{figure*}

\subsection{Bader Charge Analysis of 2D Structures/Si(111) Interfaces}
he electronic structures of the most stable interfaces of 2D material/Si(111) reveal that the group--V elemental 2D monolayers and Si surfaces are indeed metallic, irrespective of $\alpha$ and $\beta$ allotropes.  However, there are differences in the structural arrangements at/near their interface regions. Such changes arise due to the specific kind of interaction or charge distribution/transfer between Si and As, Sb or Bi atoms at their respective interfaces. In this section, we discuss these aspects through detailed Bader charge analysis with the help of charge density profiles.

The Bader charge calculations reveal that in the case of both $\alpha$-As/Si(111) and $\beta$-As/Si interfaces, the Si atoms donate a total electronic charge of 1.23$\vert$e$\vert$ and 0.56$\vert$e$\vert$ (e is the charge of an electron)  to the monolayer As atoms, respectively. This is due to the higher electronegativity of As compared to Si. However, in the case of $\beta$-Sb/Si and $\beta$-Bi/Si slabs, interfacial Sb and Bi atoms donate a total charge of 0.24$\vert$e$\vert$ and 0.31$\vert$e$\vert$ to the interfacial Si atoms, respectively. The charge transfer between Sb or Bi and Si atoms are small due to a very small electronegativity difference among them (electronegativity of Si, Sb and Bi are 1.9, 2.05, and 2.02 respectively), resulting in a strong interaction between the interfacial layer of Si and monolayers of antimonene, and bismuthene. This strong interaction leads to significant hybridization between p-orbitals of Sb, and Bi with the p-orbitals of Si, as evident from the orbital projected DOS plots (see Fig.~\ref{fig2}). Such a trend of charge transfer is also reflected in the charge density distribution plots.

Figure~\ref{fig4} displays the charge density profiles for all the heterostructures (between $\alpha$-As and $\beta$ allotrope of As/Sb/Bi and Si(111)) at/near their interfacial regions. In case of $\alpha$-As/Si and $\beta$-As/Si interfaces (Fig.~\ref{fig4}(a, b)), most of the charges are accumulated in the bonding region between Si and As atoms, confirming a covalent nature of bonding between them. The dense charge accumulation around each As atom indicates that As atoms accept charges from Si, which is well corroborated with the charge distribution results from Bader charge analysis. We have also compared the charge distributions of the interfacial arsenene monolayer with the pristine arsenene monolayer without Si. The charge density plot for pristine $\alpha$- and $\beta$-arsenene is shown in Fig.~\ref{fig4}(e, f), revealing the covalent nature of As--As bonding. A similar kind of charge distribution is also observed in interfacial As atoms in both As/Si surface slabs (see Fig.~\ref{fig4}(a, b)). Additionally, there are very small changes in bond lengths ($\sim$ 0.05 \AA\ to 0.1 \AA ) in both $\alpha$ and $\beta$ arsenene with Si interfacial structure as compared to the bond length of the pristine counterpart. This infers that both the interfaces $\alpha$ and $\beta$-arsenene sustain their pristine character and hence, the bonding feature sustains even after interacting with Si.
\begin{figure*}[t!]
	\centering
	\includegraphics[scale=0.72]{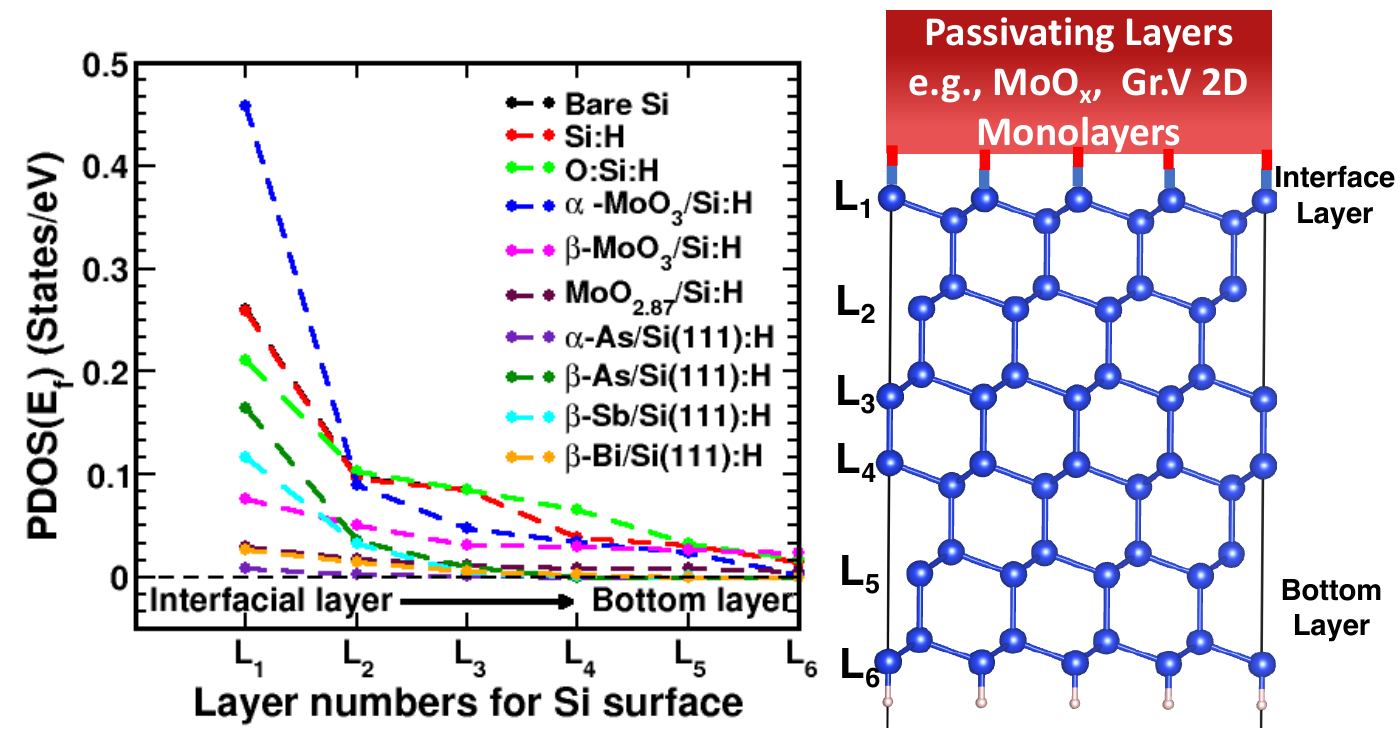}
	\caption{(Left) PDOS at E$_\mathrm{f}$ vs. the layer numbers of the Si slab for bare Si with no passivation (Bare Si), only H-passivated Si (Si:H), oxygen passivation at the top and H-passivation at the bottom Si (O:Si:H), $\alpha$--MoO$\mathrm{3}$/Si:H, $\beta$--MoO$_\mathrm{3}$/Si:H, $\beta$--MoO$_\mathrm{2.87}$/Si:H, $\alpha$--As/Si(111):H, $\beta$--As/Si(111):H, $\beta$--Sb/Si(111):H and $\beta$--Bi/Si(111):H surface slabs with the bottom Si layer passivated with hydrogen. The layer number of Si starts with 1 (at/near interface) to 6 (at bottom passivated by H). (Right) Schematic diagram of passivating layer and Si absorber interface heterojunction indicating Si layer numbers, interface layer and bottom layers of Si surface.}
	\label{fig5}
\end{figure*}

\begin{figure}[hb!]
	\centering
	\includegraphics[scale=0.45]{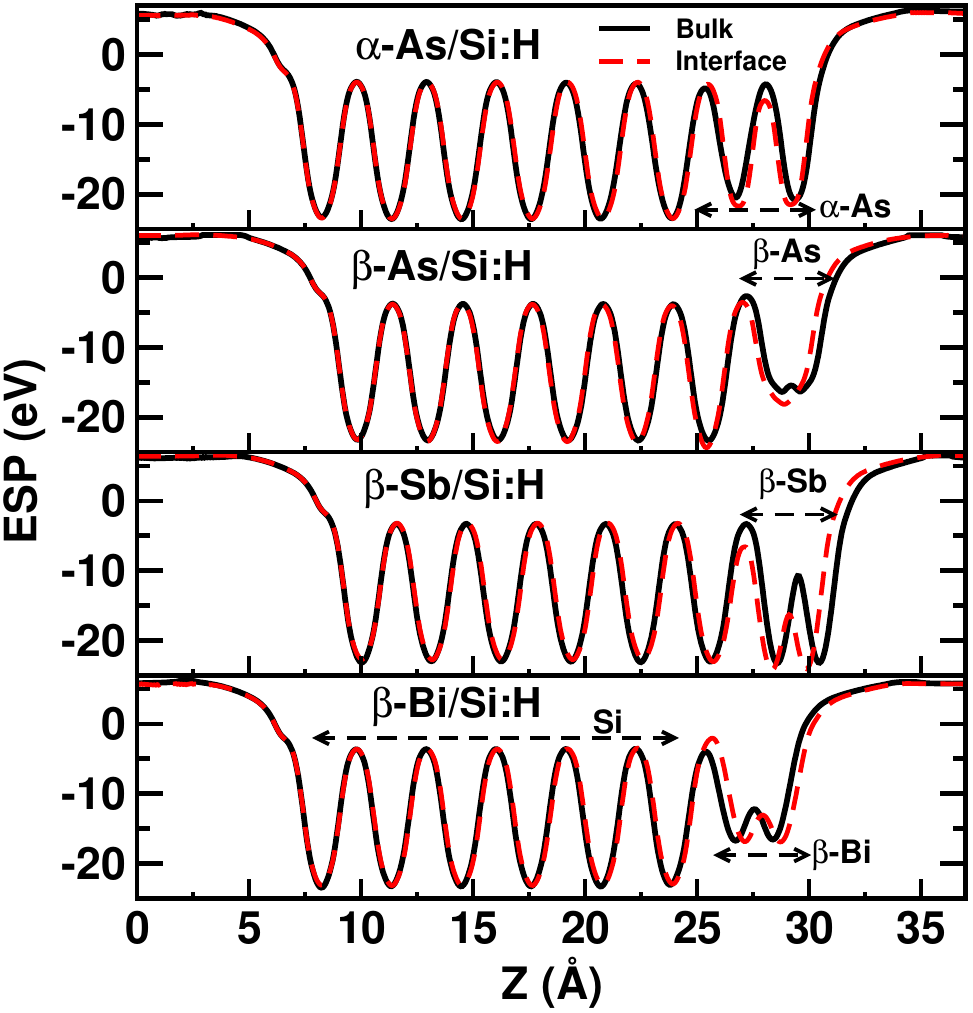}
	\caption{Planar average electrostatic potential (ESP) of $\alpha$--As/Si:H, $\beta$-As/Si:H, $\beta$-Sb/Si:H, and $\beta$-Bi/Si:H, interfaces. The solid black and red dashed line represents the average ESP for the interfacial surface slabs and their corresponding bulk counterpart, respectively.  }
	\label{fig6}
\end{figure}
Similarly, in the charge density profiles of $\beta$-Sb/Si and $\beta$-Bi/Si (see Fig~\ref{fig4}(c, d)), most of the charges are accumulated closer to the Si atoms at/near the interface. This confirms that the Si atoms act as acceptor at/near the interface, which matches our Bader charge analysis. Comparing the charge distribution in the interfacial antimonene and bismuthene layers with their pristine structures, as shown in Fig.~\ref{fig4}(g, h), it is evident that the interfacial antimonene retains its covalent character similar to pristine $\beta$-antimonene. However, in pristine $\beta$--bismuthene, the charges are localized around the Bi atoms only (see Fig.~\ref{fig4}(h)), indicating the predominant ionic bonding nature of Bi--Bi bonds. Due to the structural distortion of the bismuthene layer after interacting with Si, an abrupt charge distribution is observed between the Bi atoms (see Fig.~\ref{fig4}(d)), which is completely different from the charge distribution of pristine $\beta$--bismuthene as shown in Fig.~\ref{fig4}(h). We speculate that the interfacial Bi atoms might exhibit a less covalent and dominant ionic type metallic bonding nature. Thus, it is reasonable to expect that the impact of interfacial Si atoms are stronger to the bismuthene layers compared to arsenene and antimonene layers.

\begin{figure*}[t!]
	\centering
	\includegraphics[scale=0.55]{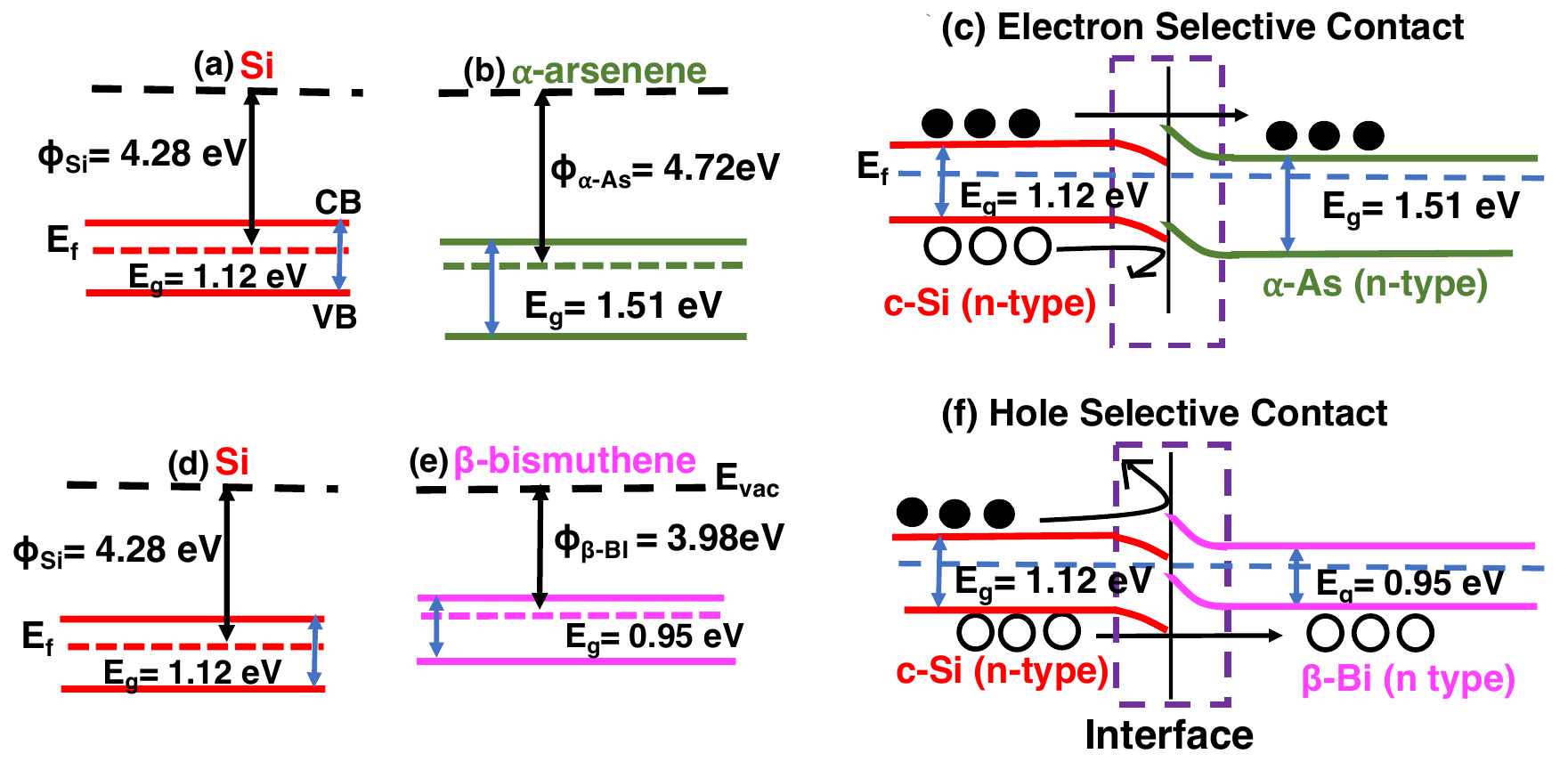}
	\caption{Flat band diagrams of (a, d) Si (top and bottom diagrams of leftmost figures), (b) $\alpha$--As (e) $\beta$--Bi, respectively. E$_\mathrm{vac}$, E$_\mathrm{f}$ and $\phi_i$ are the vacuum energy, Fermi energy and work function, respectively. VB (CB) represent the valence (conduction) band. Proposed energy band alignment for (c) monolayer $\alpha$-As/Si interface as electron selective contact and (f) monolayer $\beta$-Bi/Si interface as hole selective contact. Solid and hollow black spheres represent electrons and holes, respectively.}
	\label{fig7}
\end{figure*}

\subsection{Comaprision of surface passivation in 2D Structures/Si vs. other electron/hole passivation layer/Si interfaces}
In previous sections, we established the electronic structure of the most stable interfaces between the group--V elemental 2D materials and Si. Now, we will compare their efficiency as passivating layer to Si with respect to a few interfaces between other conventional passivating layers and Si. MoO$_\mathrm{x}$ is a well studied hole selective passivating layer in Si PV devices. For more details, please refer.\cite{behera2021first}  Figure~\ref{fig5} shows the variation of Si-PDOS at E$_\mathrm{f}$ with respect to the layer number (starting from top (interfacial layer) to bottom Si layer) for different cases \emph{i.e.,} bare Si, Si with H–passivation (Si:H), Si with both H- and O-passivation (O:Si:H), MoO$_\mathrm{x}$/Si:H and various 2D structure/Si interfacial surface slabs. For all surface structures, the contribution of electronic states decreases as we move from the top Si layer toward the bottom (bulk--like). However, the electronic states of the Si surfaces in the 2D structure/Si interfaces become negligibly small (becomes zero) after the third layer, indicating the verisimilitude behavior of bulk--like Si at the bottom layer (see Fig.~\ref{fig2}(d,h,i)).  Since all the interfacial surface slabs are metallic in nature, the relatively lower magnitude of electronic states at E$_\mathrm{f}$ is an indirect measure of better passivation of Si surface. For $\alpha$-As/Si(111):H and $\beta$-Bi/Si(111):H surface slabs, the lowest PDOS value at E$_\mathrm{f}$ as compared to other cases suggests $\alpha$-arsenene and $\beta$-bismuthene to be potential candidates as passivating layer for Si-based PV devices. It is even better than the conventional MoO$_\mathrm{x}$ system. This concludes that these 2D materials have the potential to offer better passivation as compared to other conventional dual-functional layers and hence, should be experimentally verified.

We further explored the interfacial surface reconstructions in the $\alpha$--As/Si(111):H, $\beta$--As/Si(111), $\beta$--Sb/Si(111), and  $\beta$--Bi/Si(111) interfaces by simulating their planar average electrostatic potential (ESP) of the bulk as well as the surface slabs. Figure~\ref{fig6} shows the simulated plots of the average ESP along the (001) direction of these interfacial slabs. The red dashed lines indicate the average ESP for the 2D materials/Si(111) surface slabs, while solid black lines indicate their corresponding ESP for the bulk material. As we move away from the interface region, the crystals are expected to reflect their bulk behavior.\cite{van1985theoretical} From Fig.~\ref{fig6}, it is clearly evident that the average ESP of Si for both the bulk and interfacial surface slabs of all the 2D material/Si interfaces coincides after the third layers of the Si slabs. This indicates that Si surface slab rapidly recovered its bulk characteristics after three layers in the presence of monolayers $\alpha$-As and $\beta$-As/Sb/Bi as a passivating layer. As a result, we obtain a band gap at the third layer of Si surface slabs in all the interface structures, as shown in Fig.~\ref{fig1}(d) and Fig.~\ref{fig2}(d, h, i). In the case of $\alpha$-As/Si:H interface, there is no significant shifting of average ESP observed in the interface region, as very minimal structural distortion of Si occurs in the interface after structural relaxation. Hence, the Si surface slab rapidly recovered its bulk character after three layers, and we obtain a band gap of 0.04 eV at the third layer (band gap increases with higher order layers) of Si surface slab (see Fig.~\ref{fig1}(d)). In contrast, significant lattice distortion is observed in all the $\beta$ allotropes-based interfacial Si surface slabs (see Fig.~\ref{fig2}(b, f, j)) after structural optimization. This effect is also reflected in the simulated surface average ESP, which is significantly shifted with respect to the bulk as shown in Fig.~\ref{fig6}. Therefore, the surface reconstruction in the Si surface extends over a longer ranged in the case of $\beta$ allotrope based Si interfaces. Due to high structural distortion in $\beta$--Sb/Si:H interface, the Si surface is unable to mimic its bulk character, and it remains metallic throughout the layers.  However, due to lesser structural distortion at the interface in $\beta$--As/Si:H and $\beta$--Bi/Si:H surface slabs with respect to $\beta$--Sb/Si:H surface, the average ESP of their interface and bulk coincide after the third layer and acquire a band gap of 0.6 eV, see Fig.~\ref{fig2}(d, i). As such, one can conclude that due to lower partial DOS at E$_\mathrm{f}$ and less structural distortion of Si layers at the interface region, $\alpha$--As/Si:H and $\beta$--Bi/Si:H interfaces yield better surface passivation.

\subsection{Band Alignment of 2D Structures With Si for Carrier Transport }
In Fig~\ref{fig5}, we established that $\alpha$-arsenene and $\beta$-bismuthene offer better surface passivation to Si compared to the state-of-the-art MoO$_\mathrm{x}$ layer. Further, to study the impact of charge carrier selectivity when these 2D materials are in contact with Si absorbers, it is crucial to understand their band alignment with Si. For this, we have calculated work functions ($\phi$) for $\alpha$-As and $\beta$-Bi and found to be 4.72 eV and 3.98 eV, respectively. These simulated values are in good agreement with other reported findings.\cite{jamdagni2018two} To evaluate the charge selectivity, \emph{i.e.,} whether these group-V based 2D allotropes can serve as either electron selective or hole selective layers for Si absorber in PV cells, the band alignment of $\alpha$--As and $\beta$--Bi must be analyzed with respect to Si.

Figure~\ref{fig7}(a, d, b, e) depicts the flat band diagrams of Si, $\alpha$-As, and $\beta$-Bi, respectively. These flat band diagrams are plotted considering a carrier concentration of $\sim$10$^{15}$ cm$^{-3}$ for Si (n-type), which is typically employed in commercial Si solar cells. For the 2D materials, we have considered significantly higher carrier concentrations $\approx$10$^{19}$--10$^{20}$ cm$^{-3}$, as they are potential candidates for transparent conductors.\cite{behera2021first} The vacuum energy levels shown in Fig.~\ref{fig7} for $\alpha$--As and $\beta$--are aligned with reference to the vacuum level of Si. As evident, the work functions of $\alpha$-As monolayers are higher than that of Si. As a result, the positions of their valence band maximum (VBM) and conduction band minimum (CBM) will be at lower potentials compared to those of Si. In order to understand the charge carrier transport across the interface of Si and $\alpha$-As, we have shown a proposed energy band alignment for $\alpha$-As and Si heterojunction in Fig~\ref{fig7}(c). So far, the CBM and VBM of Si are at higher potential with respect to $\alpha$-As. A downward band bending in the CBM and VBM of Si is expected at the interface (see Fig~\ref{fig7}(c)) when the heterojunction is at equilibrium. As a result, electrons can readily move from the CBM of Si towards the CBM of $\alpha$-As, and consequently, the movement of holes is hindered from the VBM of Si to the VBM of the $\alpha$-As. Thus, the band alignment analysis suggests that $\alpha$--arsenene can be a potential candidate for hole blocking electron selective contacts to Si. 

\section{Conclusion}
To conclude, we investigated the structural reconstruction and electronic properties of the group--V elemental 2D materials and Si absorber interface using first-principle calculations. The key objective is to evaluate the potential of these 2D materials as passivating electron/hole transport layers to solar PV devices.  We found that the interfacial structures of 2D layered materials on top of Si(111) surfaces exhibit greater stability as compared to Si(100) surfaces. For all interfacial configurations 2D materials/Si(111), the monolayers of $\beta$-As/Sb/Bi and Si show metallic character. In contrast, in the metastable $\alpha$-Bi/Si(100) and $\beta$-As/Si(100) interfaces, both bismuthesne and arsenene maintain their semiconducting nature, leading to providing the best surface passivation to Si. The Bader charge analysis suggests a strong interaction between Si and As, Sb, or Bi atoms due to their smaller electronegativity difference. Such strong interactions facilitate significant orbital hybridization between As-Si, Sb-Si, and Bi-Si, as evidenced by their orbital projected density of states (DOS). It is also observed that the structural distortion of the Si slabs is lowest in the case of $\alpha$--As/Si(111) and $\beta$--Bi/Si(111) interfaces compared to $\beta$--As and $\beta$--Sb based Si(111) interfaces. This leads to a lower surface density of states in the ideal band gap region of Si, and it is even lower than the well studied MoO$_\mathrm{x}$/Si interface, implying better passivation. More importantly, the energy band alignment of these two interfaces confirms that $\alpha$--arsenene can provide passivating electron selective contact and $\beta$--bismuthene can be a better candidate for incorporation as passivating hole selective contact in Si PV devices. The present study will motivate further experimental investigations on the group--V elemental 2D materials and verify their potential as electron/hole transport and passivating layers.\\

\section*{Acknowledgements}
GB acknowledges the Council of Scientific and Industrial Research (CSIR), India for providing Research Fellowship. This work was carried out as part of the National Centre for Photovoltaic Research and Education (NCPRE) Phase 2 funded by the Ministry of New and Renewable Energy (MNRE), Government of India at IIT Bombay. 
\bibliography{behera_references.bib}

\end{document}